\documentclass[a4paper, amsfonts, amssymb, amsmath, reprint, showkeys, nofootinbib, twoside]{revtex4-1}
\usepackage[english]{babel}
\usepackage[utf8]{inputenc}
\usepackage[colorinlistoftodos, color=green!40, prependcaption]{todonotes}
\usepackage{amsthm}
\usepackage{mathtools}
\usepackage{physics}
\usepackage{xcolor}
\usepackage{graphicx}
\usepackage[left=23mm,right=13mm,top=35mm,columnsep=15pt]{geometry} 
\usepackage{adjustbox}
\usepackage{placeins}
\usepackage[T1]{fontenc}
\usepackage{lipsum}
\usepackage{csquotes}
\usepackage[pdftex, pdftitle={Article}, pdfauthor={Author}]{hyperref} % For hyperlinks in the PDF
\usepackage{ulem}
\begin{document}
	\def\SB#1{\color{red!50}{#1} \color{black}}
	\def\RV#1{\color{blue!50}{#1} \color{black}}
	
	\def\RVb#1{\color{blue!50}{\sout{#1}} \color{black}}
	\def\VM#1{\color{green!50}{#1} \color{black}}
	\def\VMb#1{\color{green!50}{\sout{#1}} \color{black}}
	\def\SBb#1{\color{red!50}{\sout{#1}} \color{black}}
	
	\title{Quantitative absorption imaging of optically dense effective two-level systems}
	% samples at high saturation.}

\author{R. Veyron}
\author{V. Mancois}
\author{J-B. Gerent}
\author{G. Baclet}
\author{P. Bouyer}
\author{S. Bernon}
\email[Correspondence email address: ]{simon.bernon@institutoptique.com} % Your name
\affiliation{ LP2N, Laboratoire Photonique, Numerique et Nanosciences, Universite Bordeaux-IOGS-CNRS:UMR 5298, rue F.Mitterrand, F-33400 Talence, France}

\date{\today} % Leave empty to omit a date

\begin{abstract}
	Absorption imaging is a commonly adopted method to acquire, with high temporal resolution,  spatial information on a partially transparent object. It relies on the interference between a probe beam and the coherent response of the object. In the low saturation regime, it is well described by a Beer Lambert attenuation. In this paper we theoretically derive the absorption of a $\sigma$ polarized laser probe by an ensemble of two-level systems in any saturation regime. We experimentally demonstrate that the absorption cross section in dense $^{87}$Rb cold atom ensembles is reduced, with respect to the single particle response, by a factor proportional to the optical density $b$ of the medium. To explain this reduction, we developed a model that incorporates, in the single particle response, the incoherent electromagnetic background emitted by the surrounding ensemble. We show that it qualitatively reproduces the experimental results. Our calibration factor that has a universal dependence on optical density $b$ for $\sigma$ polarised light : {$\alpha=1.17(9)+0.255(2) b$} allows to obtain quantitative and absolute, in situ, images of dense quantum systems.
\end{abstract}


\maketitle

\onecolumngrid

%%%%%%%%%%%%%%%%%%%%%%%%%%%%%%%%%%%%%%%%%%
\section{Introduction} \label{sec:outline}

Light propagation and attenuation in a dense medium are long lasting problems that have been particularly resistant to predictive models and experiments. The formal description of light propagation in a dilute medium was first reported by Bouguer \cite{Bouguer1729} and later rediscovered by Lambert and Beer. The microscopic derivation of the Beer-Lambert law (BLL) relies on two premises: there should be no optical saturation nor the scatterers constituting the medium should influence each other. To get rid of the former constraint, the BLL can be easily modified and yields a decay of the light intensity following Lambert $W$ function \cite{Corless1996}. This solution is exact in a dilute medium of two-level atoms and was used to develop absorption imaging technics and derivatives \cite{Pappa2011}

Including the full multi-level nature of an atomic system is straightforward under perfectly polarized probing, as it is only a matter of computing the effective dipole strength \cite{Gao1993,Veyron2021}. To modify minimally the BLL, Reinaudi \textit{et al.} \cite{Reinaudi2007} originally proposed to reduce the saturation intensity with a single parameter $\alpha$, globally encapsulating any deviation from the idealized BLL. For the time being, $\alpha$ was used as a holdall that subsumes the undesired complexity arising from the presence of stray magnetic fields, imperfect probe polarization, multiple scattering or even the multilevel structure of the atomic system upon consideration. For the sake of atom number calibration, $\alpha$ has been estimated independently in numerous experiments \cite{Mordini2020,Yefsah2011a,Reinaudi2007,ReinaudiPhD,Horikoshi2017,Seroka2019,Riedel2010, Kwon2012} with very disparate results even for similar probe conditions (see Tab. \ref{tab:review_alpha}). There is no consensus \textit{ad idem} concerning the acceptable values it should take nor regarding its scaling.

% Stray magnetic fields and imperfect polarization have only little influence on $\alpha$ \cite{Veyron2021} where multilevel Optical Bloch Equations were solved in the single atom picture and obtained an effective two-level model with a well determined rescaling factor $\alpha$.

We recently demonstrated \cite{Veyron2021} by solving the multilevel Optical Bloch Equations in the single atom picture that stray magnetic fields or an imperfect probe polarization have little influence on $\alpha$, although an  incoherent pumping dramatically increases its value. 
% and incoherent pumping on the scattering/absorption cross section of $\phantom{}^{87}$Rb \cite{Veyron2021} where multilevel Optical Bloch Equations were solved in the single atom picture and obtained an effective two-level model with a well determined rescaling factor $\alpha$.

Generalizing BBL to optically dense systems appears riskier at first glance, as it is required to use saturation parameters of the order of the optical thickness \cite{Reinhard2014}. While a single atom response is exact and very well characterized experimentally in both unsaturated and saturated regimes \cite{Wrigge2007,Tey2009,Streed2012}, the response of atomic clouds involves necessarily multiple scattering and high order correlations \cite{Lee2016}, neither of them considered in the single atom model of BLL. Moreover, the geometry of the medium may favor -or hinder- a variety of many body responses under coherent illumination, as seen experimentally with endfire superradiance \cite{Inouye1999},  or as predicted theoretically in the spectroscopy of 2D arrays of atoms \cite{Bettles2020}.

It is now widely accepted that collective phenomena in dense media scale with powers of the optical thickness \cite{Labeyrie2004,Weiss2018} which is precisely the quantity that BLL endeavors to estimate. We expect in dense atomic media under resonant saturated illumination that scattered photons contribute to saturate \textit{incoherently} albeit \textit{resonantly} \cite{Pucci2017} the neighboring atoms in the forward and the backward directions. In this paper, we demonstrate experimentally that the value of $\alpha$ scales linearly with the optical density (OD). We propose a 1D model that accounts for collective effects via multiple scattering to explain this scaling.

%Calibration of the cross section using saturating absorption has been widely used to account for experimental imperfections which cause optical pumping coming from either residual probe ellipticity, stray magnetic field or transient regime. Experimentally, stray magnetic fields can be compensated by active feedback and transient regime impact can be minimized by using longer probe durations. Also, an imperfect linear or circular polarization does not play a huge role because of optical pumping which maintain the cross section towards linear or circular cases. For $^{87}$Rb, the reduction factor should range from 1 to 1.8. In Table \ref{tab:review_alpha}, we show such calibrations for $^{87}$Rb performed by different groups. We see that the value of $\alpha$ can cover values above 2 without further explanations by the authors. In this paper, we show that the value of $\alpha$ scales linearlyon the optical density. We propose a model that accounts for collective effects via multiple scattering to explain this behavior.

\begin{table}[htp]
	\begin{center}
		\begin{tabular}{|c|c|c|c|c|}
			\hline
			$\alpha$    &   b$_0$    &    Cloud type    &     Probe polarisation   &   Ref. \\
			\hline
			1.13(2)    &   0.5    &    BEC   &     Circular  &   \cite{Seroka2019}   \\
			1.11    &   1.2    &    BEC   &     Circular  &   \cite{Riedel2010}   \\
			2.0(2) & 2.5 & 1D Li condensate & Circular &  \cite{Horikoshi2017} \\
			3.15(12) & 5 & 1D BEC & Circular & \cite{Mordini2020} \\
			%3.9 & - & Sodium D2 & Circular & \cite{Kwon2012} \\
			\hline
			2.12(1)    &   4.8    &    2D MOT   &     Linear  &   \cite{Reinaudi2007}   \\
			2.6(3)    &   -    &    Quasi 2D BEC   &     -  &   \cite{Yefsah2011}   \\
			2.9   &   8.4    &    MOT   &     Linear    &   \cite{ReinaudiPhD}   \\
			\hline
		\end{tabular}
		\caption{\label{tab:review_alpha}Review of the cross-section calibrations {performed by different groups.}}
	\end{center}
\end{table}

%%%%%%%%%%%%%%%%%%%%%%%%%%%%%%%%%%%%%%%%%%
\section{\label{sec:BL_sat}Beer Lambert derivation in the saturating regime} 
In this section, we will derive the differential equation of propagation of a probe radiation (considered as a coherent field) in a continuous medium. The field at the point $\bm{r}$ is therefore obtained by summing the coherent incident field $\bm{E_c}(\bm{r})=\frac{E_{A}}{2}e^{ikz}\epsilon$ and the total scattered field obtained by the integral over space of the fields emitted by a continuous ensemble of dipoles  \cite{Chomaz2012,Tey2009} : 

\begin{equation}
	\begin{aligned}
		\bm{E}(\bm{r'}) = & \bm{E}(\bm{r}) + \iiint_{V} n(\bm{r})\frac{3E_{A}}{2k}e^{ikz}i\frac{e^{ik|\bm{r'}-\bm{r}|}}{|\bm{r'}-\bm{r}|} \left( \bm{\epsilon}-\left(\bm{\epsilon}.\bm{u_{{rr'}}}\right)\bm{u_{{rr'}}} \right) d\bm{r}^{3}
		\label{eq:coherent_field_summation}
	\end{aligned}
\end{equation}

where $\bm{u_{rr'}}=(\bm{r'}-\bm{r})/|\bm{r'}-\bm{r}|$ is a vector unit, $n(\bm{r})$ is the atomic density and $V$ is the volume of integration. For an infinite homogeneous slab of atoms of width $dz$, the integration is carried for $x,y$ in $]-\infty,+\infty[$ and $z$ in $[z,z+dz]$. 

In this expression, we consider only the dipole scattering in the far-field regime varying in $1/r$ which corresponds well to the regime of the data presented in section {\ref{sec:Setup}} ($nk^{-3}\ll1$). We also emphasize that the above expression is only valid in the weak saturation approximation when a two-level system (TLS) is well approximated by a dipole.
Performing the integration in cylindrical coordinates over a circle $R_{0}\gg1/k$ ({\textit{i.e.}} ignoring edge effects) for a constant atomic density in this disk $n(\bm{r})=n_{0}$ and a circular polarization, the first term proportional to $\bm{\epsilon}$ which is the on-axis scattering becomes $-n_{0}\frac{3E_{2}}{2k^{2}}e^{ik(z+dz)}2\pi\bm{\epsilon_{+}}dz$ and the second term depending on $\bm{u_{{rr'}}}$ which is the off-axis scattering reads $n_{0}\frac{3E_{2}}{2k^{2}}e^{ik(z+dz)}\pi\bm{\epsilon_{+}}dz$. {Eq.} (\ref{eq:coherent_field_summation}) can be simplified into:

\begin{equation}
	\begin{aligned}
		\bm{E}(z+dz) = & \bm{E}(z) \left( 1 - \frac{n_{0}}{2}\frac{6\pi}{k^{2}}e^{ikdz}dz \right) 
		\label{eq:coherent_field_summation_sigma}
	\end{aligned}
\end{equation}

which can be reformulated in a differential expression of the Beer-Lambert's law in field :

\begin{equation}
	\begin{aligned}
		\frac{d\bm{E}}{dz} = & -\bm{E} \frac{n_{0}\sigma_{0}}{2} e^{ikdz}
		\label{eq:coherent_field_summation_sigma_BL}
	\end{aligned}
\end{equation}
where $\sigma_{0}=6\pi/k^{2}$ is the absorption cross section.
For $kdz\ll1$, {Eq.} (\ref{eq:coherent_field_summation_sigma_BL}) gives the traditional exponential attenuation of the intensity:

\begin{equation}
	\begin{aligned}
		\frac{d|E|^{2}}{dz} = & -|E|^{2} n_{0} \sigma_{0} 
		\label{eq:intensity_sigma_BL}
	\end{aligned}
\end{equation}

When the medium is saturated, the amplitude of the coherent field emitted on resonance by a single TLS is reduced but its radiation pattern for a given driving polarization is unchanged. The volume integral carried in {Eq.} (\ref{eq:coherent_field_summation}) is still exact for an emitted dipole field reduced by a factor $1/(1+s_{c})$ where  $s_{c}=\frac{2\Omega_{c}^{2}}{\Gamma^{2}}$ is the saturation parameter and $\Omega_{c}$ the Rabi frequency proportional to the incident electric field amplitude. This reduction factor of the coherently emitted field is directly related to the coherent scattering rate $\Gamma|\rho_{eg}|^{2}=\frac{\Gamma}{2} \frac{s_{c}}{(1+s_{c})^{2}}$ that can be derived from the Optical-Bloch equations of a TLS. To take into account the effect of saturation, the equation of propagation in field (Eq. {(\ref{eq:coherent_field_summation_sigma})} is modified into:

\begin{equation}
	\begin{aligned}
		\bm{E}(z+dz) = & \bm{E}(z) \left( 1 - \frac{1}{(1+s_{c})} \frac{n_{0}}{2}\frac{6\pi}{k^{2}}e^{ikdz}dz \right) 
		\label{eq:field_sigma_BL_saturation}
	\end{aligned}
\end{equation}

which gives the general form of the Beer-Lambert's law for any saturation regime in intensity $I$:

\begin{equation}
	\begin{aligned}
		\frac{dI(z)}{dz} \left( 1+\frac{I(z)}{I_{\rm{sat}}} \right) = & -n_{0} \sigma_{0} I(z)
		\label{eq:intensity_sigma_BL_saturation}
	\end{aligned}
\end{equation}

where $I_{\rm{sat}}$ is the saturation intensity which is related to the cross section by $\sigma_{0}={\hbar\omega\Gamma}/({2I_{\rm{sat}}})$. For a multi-level system, it can be shown \cite{Veyron2021} that the coherent scattering rate is reduced and takes the form $\Gamma|\rho_{eg}|^{2}=\frac{\Gamma}{2} \frac{s_{c}}{(\alpha+s_{c})^{2}}$ where $\alpha$ depends on the probe polarization, residual magnetic field, detuning from the resonance or the  ambient electromagnetic background. For $^{87}$Rb, on resonance with the $|5S_{1/2}, F=2\rangle$ to $|5P_{3/2}, F=3\rangle$ cycling transition, a resonant probe with circular polarization has $\alpha=1$ (\textit{i.e.} perfect two-level system) and  $\alpha=1.829$ for linear polarization  and $\alpha \in [1,1.829]$ for any other probe polarization.  Including this correction, {Eq. (\ref{eq:intensity_sigma_BL_saturation})} becomes:

\begin{equation}
	\begin{aligned}
		\frac{dI}{dz} \left( \alpha+\frac{I}{I_{\rm{sat}}} \right) = & -n_{0} \sigma_{0} I
		\label{eq:intensity_sigma_BL_saturation_alpha}
	\end{aligned}
\end{equation}

Eq. {(\ref{eq:intensity_sigma_BL_saturation_alpha})} can be analytically integrated over the propagation direction which leads to the expression of the optical density in the saturating regime:

\begin{equation}
	\begin{aligned}
		b(x,y) = -\alpha \ln(T(x,y)) + {s_c} (1-T(x,y))
		\label{eq:OD_saturation_alpha}
	\end{aligned}
\end{equation}

where ${s_c}=I_0/I_{\rm{sat}}$ is the saturation parameter, $T(x,y)=\frac{I(x,y)}{I_0(x,y)}$ the probe transmission and $I_0(x,y)$ the incident imaging beam intensity.

{Eq.} (\ref{eq:OD_saturation_alpha}) can also be written using the Lambert function $W$ as:
\begin{equation}
	\begin{aligned}
		T(x,y)= \frac{ \alpha }{ {s_{c}} } W \left( \frac{{s_{c}}}{\alpha} e^{ \frac{{s_{c}}-b(x,y)}{\alpha} } \right)
		\label{eq:LambertW_alpha}
	\end{aligned}
\end{equation}

The optical density being an intrinsic cloud quantity, it should be independent of the probe beam properties. Following \cite{Reinaudi2007} the value of $\alpha$ in Eq. \ref{eq:OD_saturation_alpha} can be determined by minimizing the influence of the saturation intensity of the probe on the measured optical density. In the following, we show that $\alpha$ actually depends on the optical density. We then propose a model that emphasizes the role of incoherent scattering from the ensemble.

%%%%%%%%%%%%%%%%%%%%%%%%%%%%%%%%%%%%%%%%%
\section{\label{sec:Setup}Experimental setup and data}

\begin{figure}[h]
	\begin{center}
		\includegraphics[scale=0.38]{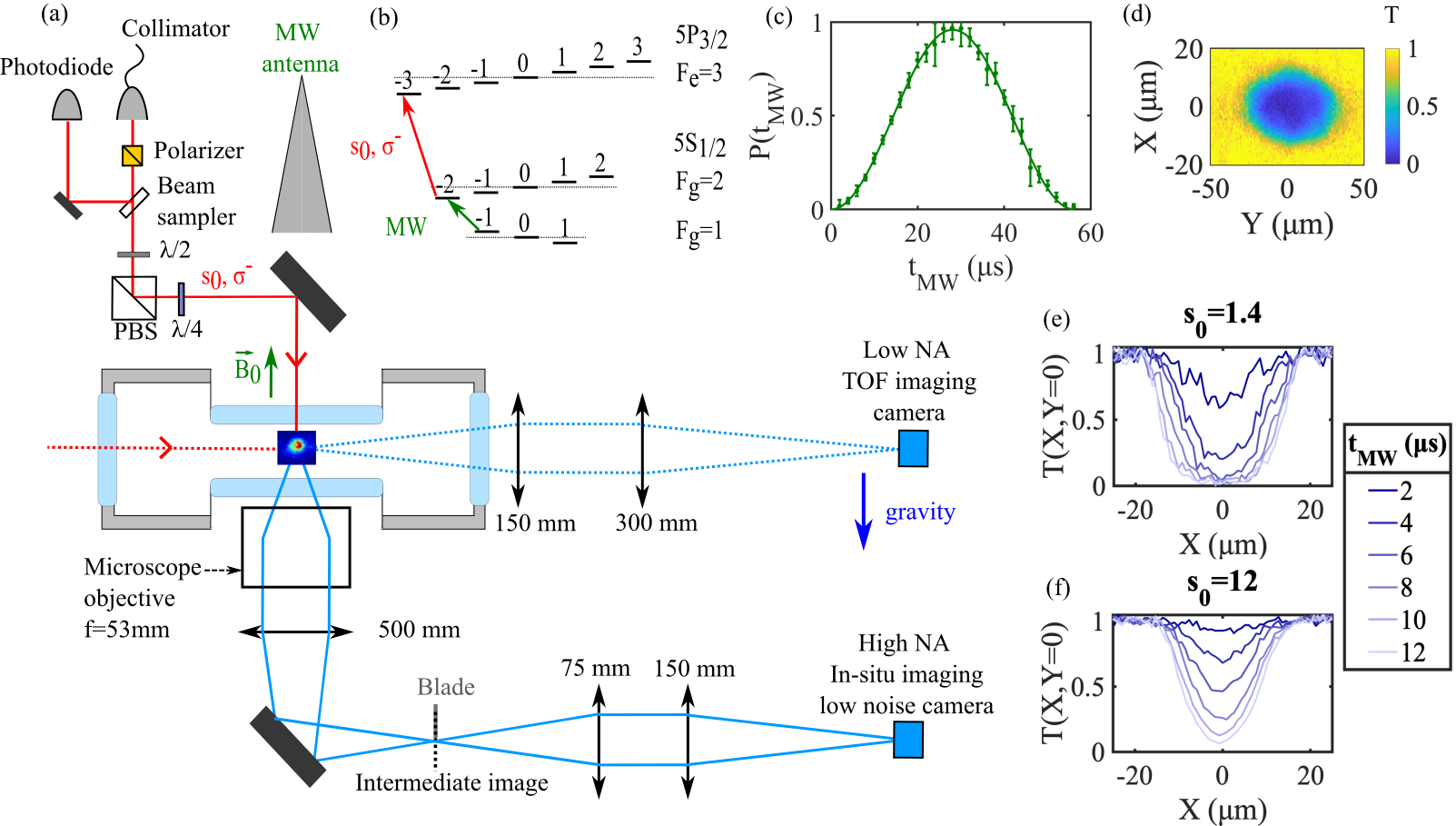}
		\caption{a) Experimental setup showing the two-imaging axis for TOF and in-situ absorption imaging. b) D2-line $^{87}$Rb atomic structure. c) Rabi oscillation of the normalized atom number (points) between $\ket{F_{g}=1,m_{-1}}$ and $\ket{F_{g}=2,m_{-2}}$ as a function of the duration and a sinusoidal fit (solid line). The data are normalized with the optically repumped total atom number. d) 2D in-situ transmission image for $s_0=1.4$ and $t_{MW}=8$ \textmu s. e)f) Cuts at $Y=0$ of the transmission for $s_0=1.3$ and $s_0=12$.}
		\label{fig:setup}
	\end{center}
\end{figure}

Thermal atoms in a pure $\ket{F_g=1,m_F=-1}$ spin state are prepared in a crossed dipole trap formed by two orthogonal and horizontal 1064 nm Gaussian beams. The dipole trap depths are $U_x,U_y={33}(3), {11}(1)$ \textmu K with respectively beam waists $w_x,w_y=50(1),65(1)$ \textmu m. The cloud has a temperature of $T=2.2(2)$ \textmu K and a total atom number of $N_{tot}=1.92(15)\times10^{5}$ which has been measured by absorption imaging of a $\pi$-polarized probe in the low saturation regime after long time-of-flight (TOF) that guarantee a peak OD lower than 0.5. The in-situ expected widths of this thermal cloud %in an harmonic trap %given by $\sigma_{i}=\sqrt{k_bT/m\omega_{i}^2}$ where $\omega_{i}$ is the trap frequency $\omega_{i}=\sqrt{4U_0/mw_{i}^2}$
are $\sigma_{x},\sigma_{y}=\sqrt{k_bT/m\omega_{i}^2}=6.5(5),14.6(9)$ \textmu m as calculated from the measured temperatures and trap frequencies ${\omega_x,\omega_y}=\sqrt{4U/mw_i^2}=2\pi.{358(18),159(8)}$ Hz. 
The ambient magnetic field at the position of the atoms are characterized by MW spectroscopy and compensated. The residual field is below 10 mG. A constant offset of $B_0=512$ mG is then applied along the imaging beam propagation axis (\textit{i.e.} gravity axis) and its direction matches a $\sigma_{-}$ configuration of the circularly polarized imaging probe.

%The atoms originating from a 2D Magneto-Optical Trap (MOT) are loaded in a 3D MOT, depumped into $\ket{F_g=1}$ and transferred in a magnetic trap where they are cooled by RF evaporation. Atoms are then transferred in the crossed dipole trap aligned below the zero of the magnetic quadrupole field. The magnetic evaporation ensure that we start with a pure $\ket{F_g=1,m_F=-1}$ spin state. 
From $\ket{F_g=1,m_F=-1}$ a control ratio of the total atom number are transferred to $\ket{F_g=2,m_F=-2}$ state by an on-resonance microwave pulse of duration $t_{MW}$. The transfer probability  $P(t_{MW})=P_{tot}\sin^2(\pi t_{MW}/T_{MW})$ has a Rabi period $T_{MW}/2=28$ \textmu s and an amplitude $P_{tot}=0.96$ (Fig. \ref{fig:setup}c). The peak atomic density is $2.1\times10^{19}$ at/m$^3$. For $t_{MW}=2$ \textmu s, we can expect a central optical density of 1.1(1) for a circular polarized probe with scattering cross section $\sigma_0$.%=2.907.10^{-13}$ m$^{2}$ is then about 1.1(1) for the shortest MW pulse that we can do of 2 \textmu s. 

After this transfer, the atoms in $\ket{F_g=2,m_F=-2}$ are imaged, in-situ, by absorption of a resonant circularly polarized imaging probe. The in-situ imaging system consists in a microscope objective (NA=0.44, $f_{eff}=53$ mm) followed by a 500 mm magnification lens that form an intermediate image. A secondary telescope magnifies this image by 2 on a low noise ($<3e^-$ rms read noise) Princeton CCD camera (16 \textmu m pixel size). For each realization, three images are acquired corresponding to the probe absorption (first image $I_{\rm{at}}(x,y)$), probe profile (second image $I_{\rm{no at}}(x,y)$) and background (third without probe $I_{at}(x,y)$). To minimise the influence of air turbulences, the consecutive images  are taken 400 \textmu s apart. A circular aperture with a 3 mm diameter is installed in the Fourier plane of the secondary telescope. It reduces the effective NA of the objective to 0.185 and increases the depth of field to {22} \textmu m, larger than the cloud width. The probe outcoupled from a single mode fiber has a measured waist of $w=1.13$ mm. Using the intermediate imaging plane, we measured a radial positioning offset to the beam center of 464 \textmu m between the imaging probe Gaussian profile center and the center of the cloud leading to a reduction factor of the used intensity on the atoms by $0.71$ with respect to the probe center. 

% using formula from PhD: for aperture of D'max=3 mm diameter

To make quantitative comparison between each experimental realization, the preparation of the cloud in $\ket{F_g=1,m_F=-1}$ is kept constant for all data sets. Only the MW duration ($t_{MW}=[2,4,6,8,10,12]$ \textmu s) and probe saturation (${s_c}=[0.44, 0.63,1.36, 2.2, 4.9, 7.8,12.3,16.3, 20.9, 28, 37.3, 49]$) are varied. By adjusting the probe pulse duration from 12.9 \textmu s to 3.7 \textmu s, the number of scattered photons per atom is kept constant at $\approx 70$. In absence of atoms and taken into account the imaging system transmission (T={0.76}), each pixel receives in between {200} and {6500} photons depending on the probe saturation. Each couple of parameters [$ t_{MW}, s_0$] is repeated  5 times for averaging. 

% 1.6e-3*(16e-4)^2/(500/53*2)^2/(1.05e-34*2*pi*3e8/780e-9)*22e-6*0.38 = 379
% 1.6e-3*(16e-4)^2/(500/53*2)^2/(1.05e-34*2*pi*3e8/780e-9)*3.7e-6*49 = 8220

% including transmission loss and taking the second set of parameters and Isat=16.7 W/m2:
% (0.44,12.9us): 0.44*12.9e-6*16.7*(16e-6/(500/53*2))^2/6.62e-34/(3e8/780.24e-9)*0.76
% (49,3.7us): 49*3.7e-6*16.7*(16e-6/(500/53*2))^2/6.62e-34/(3e8/780.24e-9)*0.76

% scattered photon per a tom for all runs by adjusting the imaging pulse duration from 22 \textmu s to 3.7 \textmu s.
To compensate for shot-to-shot fluctuations, the imaging pulses  are all acquired by a calibrated photodiode from which the saturation intensity of each image are independently computed. % atoms by doing a knife-edge measurement at the intermediate image plane. Finally, the beam waist at the position of the atoms is measured by deviating the beam towards a beam profiler which gives a waist at the atom position of $w=1.13$ mm. 
%For each experimental realization [$t_{MW}, s_0$], three images are acquired
%The data consists in varying the saturation parameter of a resonant $\sigma^-$-polarized imaging beam for many MW durations. Three images are taken in each experimental realization. for each run: the first one with atoms, the second one without atoms and the last one for the background. A dead time between the images of 600 ns is let for the fast kinetics image transfer. 
We now analyze the local transmission of each cloud at the position $(x,y)$ as computed by $T=\left(I_{at}-I_{back}\right)/(I_{no,at}-I_{back})$.

\FloatBarrier
%%%%%%%%%%%%%%%%%%%%%%%%%%%%%%%%%%%%%%%%%
\section{\label{sec:Setup}Data}

\begin{figure}
	\begin{center}
		\includegraphics[scale=0.18]{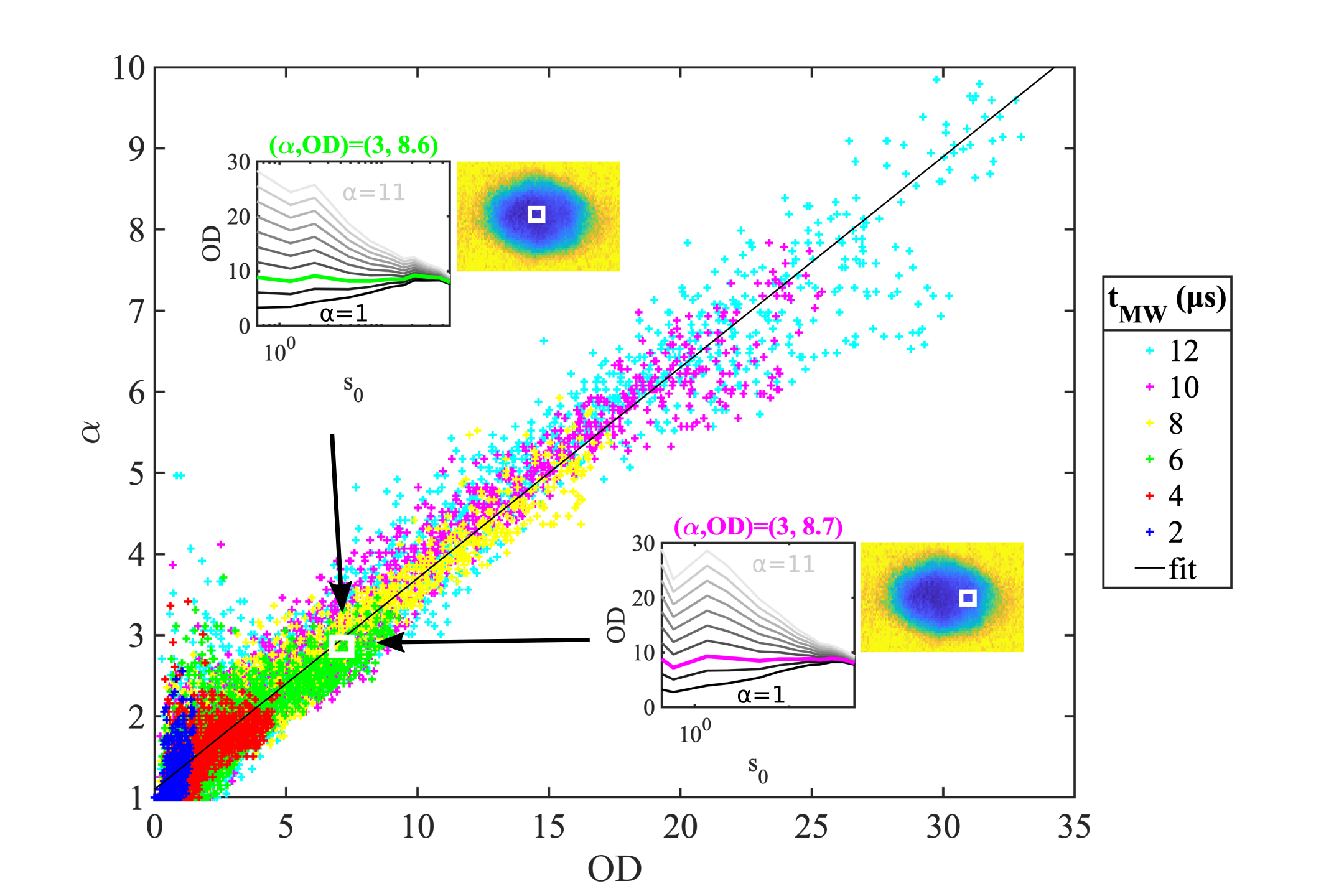}
		\caption{Correction factor $\alpha$ vs. the optical density for all MW datasets ($2, 4, 6, 8, 10, 12 $ \textmu s). Each point corresponds to 1 pixel and the solid line is a linear fit to the data {$\alpha=1.17(9)+0.255(2) b$}.}
		\label{fig:DGO_analysis_OD_alpha}
	\end{center}
\end{figure}

Following the spirit of \cite{Reinaudi2007}, for each MW duration we use the transmission acquired for the various saturation intensity and compute, for every pixel $(x,y)$, the couple of parameter ${b(x,y),\alpha(x,y)}$ that make the OD calculated from {Eq. (\ref{eq:OD_saturation_alpha})} independent on the probe saturation. The resulting parameters are plotted in {Fig.} \ref{fig:DGO_analysis_OD_alpha} where we can observe a clear correlation. 

%For each pixel of a given image, we compute the optical density by varying in the analysis the value of $\alpha$ so we get an optical density which depends on $\alpha$: 
%\begin{equation}
	%\begin{aligned}
		%b(x,y,\alpha) = -\alpha\ln\left(T(x,y)\right) + s_0\left(1-T(x,y)\right)
		%\label{eq:OD_GO_alpha}
		%\end{aligned}
	%\end{equation}

In the insets, we show 10 curves of $b$ vs. ${s_c}$ corresponding to $\alpha$ varying from 1 (black) to 11 (light grey).  The best value of $\alpha$ minimizes std($b$). To reject the noise at very low transmission that is influenced by camera read noise and fluorescence we limit the analysis to transmissions T in the range $[0.05,+\infty[$. %This range rejects noises for very low transmission due to readout noise and fluorescence and keep noises for transmissions about 1 such that noises averages. 
From the value of $\alpha$, the optical density is obtained by averaging it over all values of $s_0$. The upper inset corresponds to a pixel at the center of the cloud ($(b_c,\alpha_c)=(8.5,2.9)$) for $t_{MW}=6$ \textmu s and the lower inset to a side shifted pixel ($(b_s,\alpha_s)=(8.3,3)$) with $t_{MW}=10$ \textmu s. As the difference of position of the pixels compensates for the difference of central densities of the clouds, these two pixels correspond to a similar local optical density.  Independently from their difference of position in the cloud, equivalent local optical densities lead to the same reduction of the scattering cross section. A linear fit of the entire dataset in Fig. \ref{fig:DGO_analysis_OD_alpha} gives a slope of {$0.255(2)$} an offset of {$1.17(9)$ where the uncertainty is dominated by the uncertainty of the saturation parameter.} This offset close to 1 shows that in the limit of low densities the atomic response is well modeled by an ensemble of independent TLS. The value of $\alpha$ at low atomic density depends both on the probe polarization and magnetic field direction but also linearly depends on the calibration of the saturation intensity as observed by the atomic cloud. The calibration of the offset between probe center and atoms was important in this respect. The dependance of $\alpha$ on $b$ shows that $\alpha$ is not solely determined by the probe properties but also depends on the optical density which is a signature of the influence of multiple scattering.
%We show in Fig. \ref{fig:DGO_analysis_OD_alpha} for instance curves of $b(\alpha)$ for two different pixels. 
%At the center of the cloud, for a MW pulse of $6$ \textmu s, we clearly see that taking $\alpha=1$ would not make the optical density independent of the saturation parameter. So the correct value is rather $\alpha=2.9$ and $b=8.5$ which gives the green curve. Interestingly, a pixel shifted on the side of the cloud for a longer MW pulse of $10$ \textmu s gives $\alpha=3$ and $b=8.3$ as the magenta curve is flat with  these parameters. This means that 
%All MW curves converge towards $\alpha=1$ at low optical densities. In this case, the value is dependent on experimental imperfections like the precise calibration of the saturation parameter.

%We fit the curve with a linear function and extract a slope of $0.26$ and an offset of $1.1$. 

\FloatBarrier
%%%%%%%%%%%%%%%%%%%%%%%%%%%%%%%%%%%%%%%%%
\section{\label{sec:Setup}Model}

As shown in \cite{Veyron2021}, the coherent scattering properties of a TLS probed by circularly polarized light are almost insensitive at first order to experimental imperfections such as magnetic bias orientation or incident polarization. Nevertheless, in the same paper, it was  shown that the presence of an incoherent electromagnetic background will influence the coherent response of atoms. Here, we show that such incoherent background could arise from multiple scattering in the medium and we propose a model of propagation that can be quantitatively compared to the data.  In the model, we consider the propagation of the coherent probe in a cloud that is homogeneous in the transverse directions and has Gaussian density profile in the propagation direction. The system is translation invariant in the transverse plane $(x,y)$. The electromagnetic field and intensity therefore only depend on $z$. For a resonant $\sigma$ polarized probe, the scattering properties of an atom embedded in an incoherent electromagnetic background are well described \cite{Veyron2021} by an effective TLS. Its  coherent and total scattering rates, as deduced from the density matrix calculation, are given by  :
\begin{equation}
	\begin{aligned}
		R^{(coh)}_{sca} = & \frac{\Gamma}{2\alpha} \frac{{\frac{s_{c}}{\alpha}}}{\left( 1+\frac{s_{c}}{\alpha} \right)^{2}} \\
		R^{(tot)}_{sca} = & \frac{\Gamma}{2} \left( \frac{\frac{s_{c}}{\alpha}}{1+\frac{s_{c}}{\alpha}} +\frac{\frac{s_{i}}{\alpha_c}}{1+\frac{s_{i}}{\alpha_c}}\right)
		\label{eq:RhoeeRhoegSteady_Coherent_Incoherent_2level}
	\end{aligned}
\end{equation}

where the corrected scattering rates are $\alpha = \alpha_{\rm{SA}} \left(1+s_{i}\right)$ and $\alpha_c = 1+s_{c} $, where $\alpha_{SA}$ is the single atom reduction of the cross section that  depends on polarization, detuning and offset field. In this work, we have $\alpha_{SA}=1$ corresponding to on resonance $\sigma$ polarization.

\begin{figure}[h]
	\begin{center}
		\includegraphics[scale=0.18]{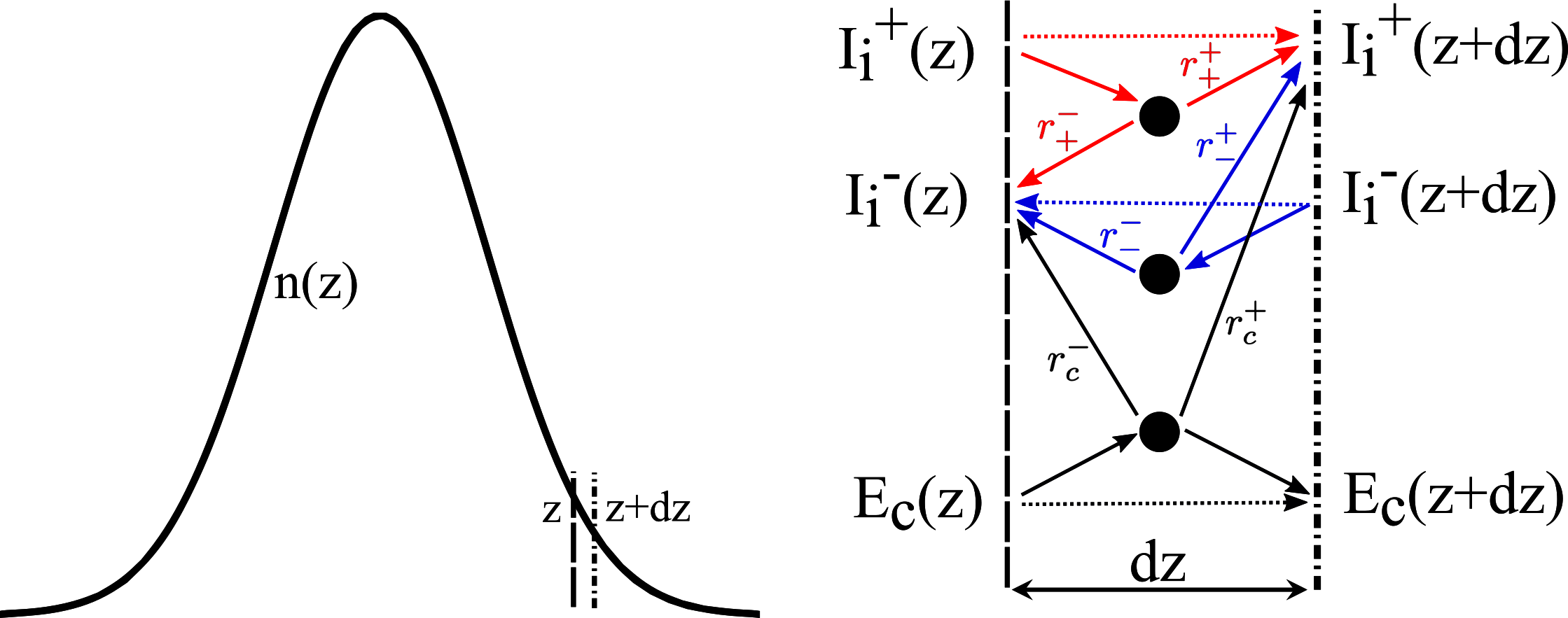}
		\caption{1D  model for the coherent and incoherent propagation and conversion of light between adjacent layers. The coherent and incoherent scattering rates are given by eq. \ref{eq:RhoeeRhoegSteady_Coherent_Incoherent_2level}. The spatial repartition are taken as $r_{+}^{+}=1-r_{+}^{-}=r_{-}^{-}=1-r_{-}^{+}=r_{c}^{+}=1-r_{c}^{-}=1/2$ (see text).}
		\label{fig:layer_dx_diagram_coh_inc_sca}
	\end{center}
\end{figure}

To numerically solve our model, we divide the propagation direction in infinitely small slabs of width $dz$. For each slab, we derive from the scattering rate expressions (Eq. \ref{}) a set of differential equations that relates  the intensity profile of the coherent intensity $I_c(z)$, forward incoherent intensity $I_i^+(z)$ and backward incoherent intensity $I_i^-(z)$. The set of equation are expressed in terms of the nondimensionalized saturation intensity parameters $s_{c}(z)=I_c(z)/I_{\rm{sat}}$, $s_{i}^+(z)=I_i^+(z)/I_{\rm{sat}}^{\rm{iso}}$, $s_{i}^-(z)=I_i^-(z)/I_{\rm{sat}}^{\rm{iso}}$ where $I_{\rm{sat}}^{\rm{iso}}=\alpha^{\rm{iso}} I_{\rm{sat}}=2.12 I_{\rm{sat}}$.
\begin{equation}
	\begin{aligned}
		\frac{ds_{c}}{dz} = & -  n(z) \sigma_{0} \frac{s_{c}}{1+s_{c}+s_{i}}  \\
		\frac{ds_{i}^{(+)}}{dz} = & -  \frac{n(z) \sigma_{0}^{\rm{iso}}}{2} \left(  \frac{s_{i}^{(+)}-s_{i}^{(-)}}{1+s_{c}+s_{i}}    -    \frac{s_{c}(s_{c}+s_{i})}{\left( 1+s_{c}+s_{i}\right)^{2}}  -    \frac{s_{c}}{\left( 1+s_{c}+s_{i}\right)^{2}}  \right)  \\
		\frac{ds_{i}^{(-)}}{dz} = & -  \frac{n(z) \sigma_{0}^{\rm{iso}}}{2} \left( \frac{s_{i}^{(-)}-s_{i}^{(+)}}{1+s_{c}+s_{i}}    -    \frac{s_{c}(s_{c}+s_{i})}{\left( 1+s_{c}+s_{i}\right)^{2}}  -   \frac{s_{c}}{\left( 1+s_{c}+s_{i}\right)^{2}}   \right)
		\label{eq:BL_coupled_coh_incoh_saturations}
	\end{aligned}
\end{equation}

where the isotropic cross section is $\sigma_{0}^{\rm{iso}}={\hbar\omega\Gamma}/({2I_{\rm{sat}}^{\rm{iso}}})$ and the total incoherent intensity is $s_{i}=s_{i}^{(+)}+s_{i}^{(-)}$. The prefactor $1/2$ corresponding to $r_{c/+/-}^{+/-}$ in Fig. \ref{fig:layer_dx_diagram_coh_inc_sca} accounts for equally distributed of backward and forward scattered intensities. In the derivatives of $s_{i}^{\pm}$ in Eq. (\ref{eq:BL_coupled_coh_incoh_saturations}), the first term account for re-scattering of incoherent field which is isotrope ($r_{+}^{+}=r_{+}^{-}$ and $r_{-}^{-}=r_{-}^{+}$), the second one accounts for the temporally incoherent ($ti$) scattering of the coherent field (\textit{i.e. resonant fluorescence}) which is also isotropic ($r_{c}^{-,ti}=r_{c}^{+,ti}$). The last term that conserves energy corresponds to the coherent field scattering in a temporally coherent but spatially incoherent ($si$) field arising from the discrete random position of atoms in an ensemble that we assume to be isotropic ($r_{c}^{-,si}=r_{c}^{+,si}$). This spatially incoherent field is out-of-phase with the coherent probe and its effect on the coherence term of the density matrix $\rho_{eg}$ will spatially average to 0. As checked numerically, for large saturation, the scattering being mostly temporally incoherent, this assumption has very little consequence on the results. In this model both temporally and spatially incoherent contributions are summed in the incoherent intensity : $r_{c}^{+/-}=r_{c}^{+/-,ti}+r_{c}^{+/-,si}$.
%Even though its physical interpretation is different will be out-of-phase with the but  scattering of  that redistribute third term on the right hand side represent the coherent scattering that is not forward. This term makes the energy conserved when all equations in (\ref{eq:BL_coupled_coh_incoh_saturations}) are summed up.

\begin{figure}[h]
	\begin{center}
		\includegraphics[scale=0.15]{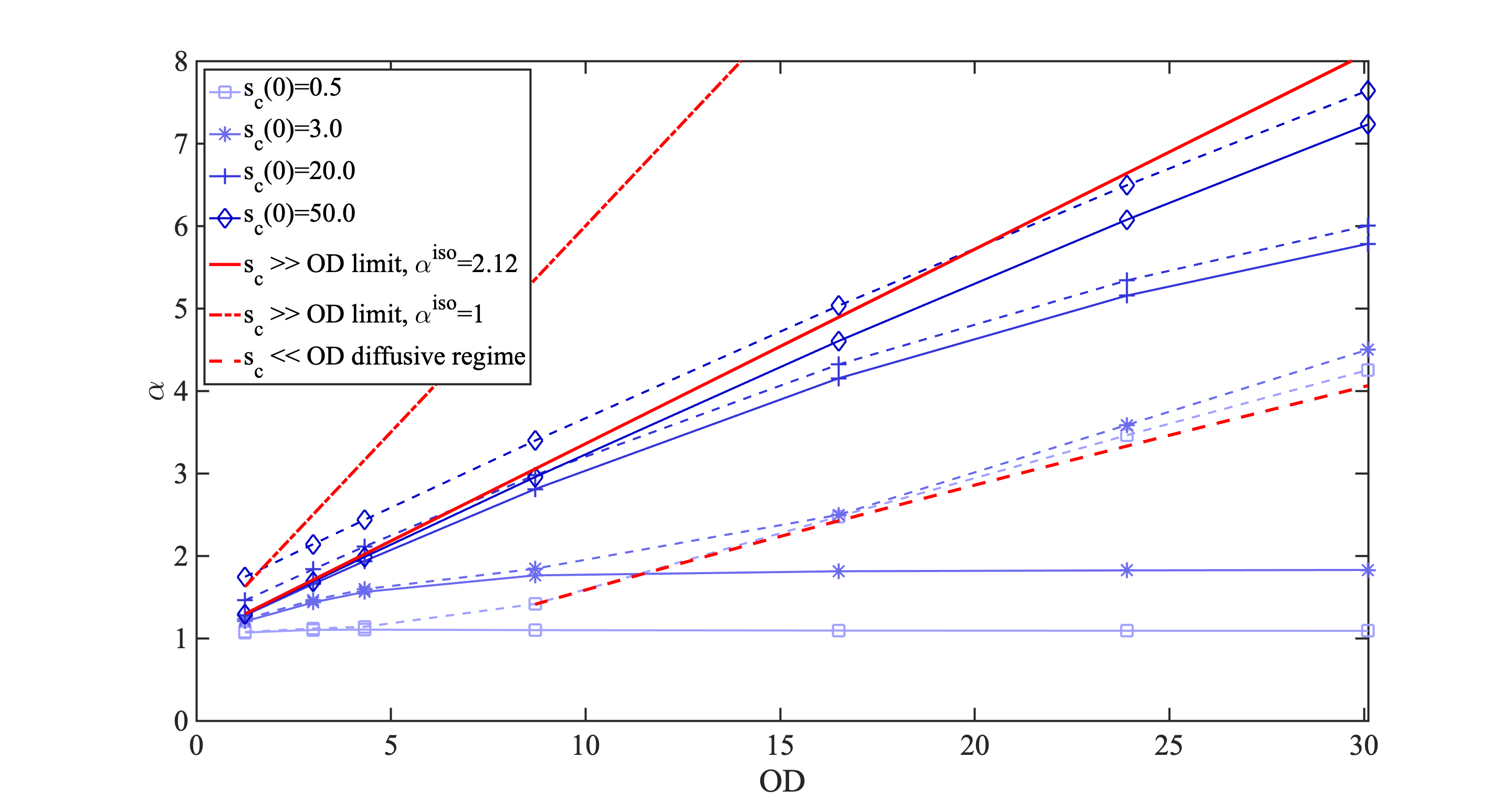}
		\caption{$\alpha$ parameters calculated from the saturated scattering model (Eq. \ref{eq:BL_coupled_coh_incoh_saturations}). The simulated probe saturation intensities are comparable to the experiment : $s_c=0.5$ in light blue (bottom) to $s_c=50$ in dark blue (top). The line (resp. dashed line) correspond to the value of $\alpha$ as calculated from the sole coherent transmission $T_c=s_c(L)/s_c(0)$ (resp. the total transmission $T=(s_c(L)+\alpha^{\rm{iso}}s_i^+(L)\Omega/(2\pi))/s_c(0)$, where $\alpha^{\rm{iso}}s_i^+(L)\Omega$ is the fluorescence background emitted in the solid angle $\Omega$ of the imaging system). In red, the analytical large saturation limit for (plain line) an effective TLS approximation of a MLS with $\sigma_0^{\rm{iso}}=\sigma_0/\alpha^{\rm{iso}}$, (dot-dashed line) an ideal scalar TLS corresponding to $\sigma_0^{\rm{iso}}=\sigma_0$ and (dashed) the diffusive regime (see text).}
		\label{fig:model_OD_alpha}
	\end{center}
\end{figure}

The value of $\alpha$ for different probe intensities are presented in {Fig.} \ref{fig:model_OD_alpha} as a function of the integrated optical density. For small probe saturation, the conversion of the coherent probe field into incoherent intensity ($s_i$) cannot generate high incoherent intensity ($s_i\ll1$) and should therefore not affect the value of $\alpha$ in the coherent propagation (Eq. {(\ref{eq:BL_coupled_coh_incoh_saturations})}). Nevertheless, for high optical density, we are in the diffusive regime. Along the propagation, while the coherent field is exponentially reduced in the mean free path length $l_{sc}=1/(n\sigma_0)$ and quickly disappears, the incoherent field, that will also be detected on the camera, is only algebraically reduced $T_{diff}\propto1/b=C/b$ \cite{Garcia1992,Guerin2017}. In this regime, it is the dominating contribution that scales as $\alpha\approx b/\ln(2\pi b/C \Omega)$ where $\Omega$ is the solid angle of the imaging system (dashed red asymptote in Fig. \ref{fig:model_OD_alpha} with $C=1$). Here ($b\gg s_c$), the increase of $\alpha$ does not correspond to a reduction of the absorption cross section but rather to an excess of detected light. In the opposite high saturation regime ($s_c\gg b$), the probe intensity is little depleted, and the incoherent saturation intensity becomes homogeneous along the propagation. By energy conservation we have $s_i={s_c}(1-T)/(2\alpha^{\rm{iso}})$. For high saturation, the last term in eq. \ref{eq:OD_saturation_alpha} dominates in the expression of the optical density $b\approx s_0(1-T)$. It leads to an expected reduction of the coherent absorption cross section that scales as $\alpha=1+s_i \approx 1+b/(2\alpha^{\rm{iso}})$ (red curve in {Fig.} \ref{fig:model_OD_alpha}). Fig. \ref{fig:model_OD_alpha} shows the expected value of $\alpha$ for a scalar TLS model where both coherent and incoherent scattering are considered $\sigma$ polarized ($\alpha^{\rm{iso}}=1$). We observe that it over-evaluates $\alpha$ while an effective TLS with isotropic incoherent scattering $\alpha^{\rm{iso}}=2.12$ has a slope 1/(2*2.12)=0.24 in good correspondence with the experimentally measured one (0.26). Given that the experimental value of $\alpha$ is averaged over all $s_c$ for a given $b$ and that this model does not take into account the transverse inhomogeneity of the cloud or the Mollow spectrum of the incoherently scattered light, the so close correspondence of the slopes is certainly fortuitous. We nevertheless believe that this model captures most of the physical origin for the increase of $\alpha$ with $b$.

\section{Conclusions} \label{sec:conc}

In conclusion we have shown that the reduction of the apparent absorption cross section is connected, in the diffusive regime ($b\gg s_c$), to the residual diffuse transmission and, in the saturating regime, to the ambient electromagnetic background originating from multiple incoherent scattering in the cloud ($s_i$). In both cases, this correction is shown to  scale mainly linearly with the optical density and reproduces well the experimentally observed dependance of $\alpha$.  In contrast with previously proposed calibration methods \cite{Reinaudi2007} that are commonly used in the community, this study shows that the reduction of cross section is not unique for an entire cloud. For a circular probe polarization under well controlled  magnetic field orientation and laser detuning, the correction factor : {$\alpha=1.17(9)+0.255(2) b$} seems universal and independent of the position in the cloud. {The offset $\alpha_0=1.17(9)$ ultimately depends on a fine calibration of the saturation parameter and should be equal to 1 for a perfectly calibrated $\sigma$ light.} A similar calibration could certainly be performed for other typical configuration such as linear $\pi$ probes. A strength of the proposed model is to take into account both the saturated response of a single atom embedded in an electromagnetic environment and the collective participation of the surrounding atoms to this environment in a self-consistent solution. At the cost of numerical computation power, the proposed 1D model could certainly be extended to 3D and incorporate modifications on the reabsorption cross section.

%and cross should be modified 
%and that we cannot determine a single $\alpha$ for the entire cloud as proposed in \cite{Reinaudi2007}. 
%comment utiliser la m\UTF{00E9}thode : pas du point par point pour kes petites OD\\
%Remarque sur les hypoth\UTF{00E8}ses d'homog\UTF{00E9}nit\UTF{00E9} et propagation transverse

%\twocolumngrid

\bibliography{QI_bibVarxiv}

%\appendix*
%\input{sections/appendix1.tex}

\end{document}